\documentclass[a4paper]{article}

\usepackage{graphicx,float,stfloats,subcaption}
\usepackage{calc,multirow,multicol,booktabs,titlesec}
\usepackage{amssymb,amsfonts,amstext,amsmath,amsthm}
\usepackage{algorithm,algorithmic,titlesec,verbatim}
\usepackage{pslatex,epsfig,apalike,hyperref}
\usepackage[bottom]{footmisc}
\usepackage[utf8]{inputenc}

\begin{document}

\title{\LARGE SQLi Detection with ML: \\A data-source perspective}

\author{\textbf{Balazs Pejo} and \textbf{Nikolett Kapui}\\
ELKH-BME Information Systems Research Group, \\
Laboratory of Cryptography and System Security, \\
Department of Networked Systems and Services, \\
Faculty of Electrical Engineering and Informatics, \\
Budapest University of Technology and Economics, \\
Műegyetem rkp. 3., H-1111 Budapest, Hungary\\
\textit{pejo@crysys.hu}, \textit{kapui.niki@gmail.com}}

\date{}
\maketitle

\textbf{Abstract.} Almost 50 years after the invention of SQL, injection attacks are still top-tier vulnerabilities of today's ICT systems. Consequently, SQLi detection is still an active area of research, where the most recent works incorporate machine learning techniques into the proposed solutions. In this work, we highlight the shortcomings of the previous ML-based results focusing on four aspects: the evaluation methods, the optimization of the model parameters, the distribution of utilized datasets, and the feature selection. Since no single work explored all of these aspects satisfactorily, we fill this gap and provide an in-depth and comprehensive empirical analysis. Moreover, we cross-validate the trained models by using data from other distributions. This aspect of ML models (trained for SQLi detection) was never studied. Yet, the sensitivity of the model's performance to this is crucial for any real-life deployment. Finally, we validate our findings on a real-world industrial SQLi dataset.

\section{Introduction}

One of the biggest security concerns today is Structured Query Language Injection (SQLi), which is also reflected in the OWASP Top 10 List: in 2021, injection attack was the third most common security flaw/vulnerability developers need to protect their applications from \cite{OWASP}. Furthermore, not only the occurrence but the complexity and the severity are increasing of the SQLi cases\footnote{\url{https://www.informationisbeautiful.net/visualizations/worlds-biggest-data-breaches-hacks}}, so faster and easier methods are needed to tackle this problem. Yet, due to the manifold nature of this issue, it is not easy to develop a comprehensive solution. Following the recent success of Machine Learning (ML) in other complex fields such as computer vision (CV) and natural language processing (NLP), traditional SQLi detection techniques are also being challenged by ML techniques \cite{jemal2020sql}.

In ML, the main idea is to let the algorithm detect patterns in the data to make decisions, i.e., the decisions are based on the data and not explicitly programmed into the algorithm. In general, the larger the dataset used for training, the better the model's prediction. Yet, as we will show later, most existing works only consider small datasets or use a single source. This is a severe overlook in the era of federated learning \cite{li2020federated}, where multiple entities train a single model in a semi-privacy-preserving manner.

\begin{figure}[!b]
    \centering
    \includegraphics[width=5.6cm]{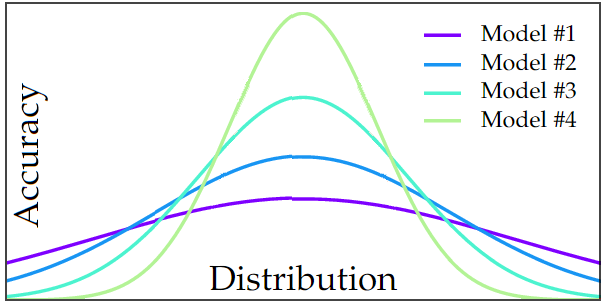}
    \caption{\footnotesize Illustrating the relationship between model accuracy and dataset distributions: the best model trained on a specific distribution could be outperformed on other distributions.}
    \label{fig:ill}
\end{figure}

To prevent overfitting on a specific distribution, it is essential to use multiple sources for training. Furthermore, to provide a comprehensive view of the trained model's performance (i.e., to indicate the model's real-world applicability), it is also vital to use multiple sources for testing. An example is provided in Figure \ref{fig:ill}: Model \#3 is less accurate than Model \#4 on distributions similar to the training distribution (at the middle of $x$ axis), while it does perform better on more distinct distributions (toward the edge of $x$ axis). Although crucial, we are unaware of any works verifying the trained models on datasets from other distributions in the SQLi context. 

Another aspect where most previous works fall short is the model evaluation: using one metric (e.g., accuracy) to measure the model's performance is insufficient, as it does not consider Type I and II errors separately. A false negative error could have a more significant impact than a false positive error in the security domain, such as SQLi detection. Although these are encapsulated in the recall and the precision, the whole picture is only captured via the Receiver Operating Characteristic (ROC) curve. It aids in optimizing the trade-off between these errors to reduce the false-negative rates (FNR) sufficiently while keeping the number of false positives rates (FPR) manageable. Such information is imperative for Security Operation Center (SOC) operators working with Security Information and Event Management (SIEM) systems. 

\vspace{0.1cm}
\noindent\textbf{Contribution. }
In this work, we highlight the shortcomings of the previous ML-based results focusing on 1) the evaluation methods, 2) the optimization of the model parameters, 3) the distribution of utilized datasets, and 4) the feature selection. Since none of the previous works explored these aspects in depth, we fill this gap, i.e., we compare different types of ML algorithms (e.g., Logistic Regression, Support Vector Machine, Random Forest, Gradient Boosting, and Neural Network) with various pre-processing methods (e.g., TF-IDF Vectorizer, Keyword weights, and Skip-gram model), and train several instances of them using hyper-parameter optimization. Additionally, we cross-verified the models on datasets corresponding to different distributions than the training samples. We also validate our findings on a private SQLi dataset originating from a major player in the security industry in Europe. 

Our findings revealed that the model with the highest accuracy is not necessarily the best choice 1) when a specific (e.g., low) false positive rate is desired and 2) when the model is used on data from other distributions. Our goal is to raise awareness of the issues using pre-trained off-the-shelf ML models and to ease the choice of security engineers in selecting the proper setup for specific use cases. 

\vspace{0.1cm}
\noindent\textbf{Organization. }
In Section \ref{sec:2}, we give a high-level introduction to both ML and SQLi.
In Section \ref{sec:3}, we provide a short taxonomy of the research efforts with an explicit focus on the issues such as evaluation methods, optimization of the model parameters, distribution of utilized datasets, and feature selection.
In Section \ref{sec:4}, besides discussing our experimental settings, i.e., the utilized datasets and explored hyper-parameters, we also detail the results of the three experiments we studied.  
Finally, we conclude our work in Section \ref{sec:5}. 

\section{Prelimilaries}
\label{sec:2}

In this section, we introduce SQLi and detail how ML is related to the issue of SQLi detection.

\subsection{SQL Injection}

SQL is a query language for relational databases to help modify, retrieve, and store data. There are many dialects, such as MySQL, PostgreSQL, and SQLite. SQL Injection is a server-side attack where a web security vulnerability allows attackers to alter the SQL queries made to the central database; therefore, they can retrieve information from or about the database, which often comes with the leakage of sensitive data. There are several types of it, as we detail below. 

\vspace{0.1cm}
\noindent\textbf{In-band SQLi. }
When the attacker could use the same channel for attacking and receiving results, an SQLi could happen by appending the results to the original query. The attack could be Error-based and Union-based, where the former tries to retrieve information about the database structure from the error messages, and the latter combines the results of two separate SELECT queries with a UNION operator. 

\vspace{0.1cm}
\noindent\textbf{Out-of-band SQLi. }
When the response does not return the SQL query result on the same channel, it is still possible to retrieve information if certain features are enabled on the site, like HTTP, DNS, or FTP (which is the case with all popular SQL servers). In this case, the attack commands the application to send data to a remote endpoint they control.

\vspace{0.1cm}
\noindent\textbf{Blind SQLi. }
When the response does not return the SQL query result, the attacker could still probe the server and observe how it behaves. The attack could be Content-based and Time-based, where the former sends conditional statements and analyzes the response, and the latter tries to make the database wait based on a condition that could be detected. 

\subsection{ML techniques}

A key technique to tackle SQLi is to reduce the attack surface by avoiding dynamic SQL statements, where the attacker cannot execute arbitrary SQL commands. Unfortunately, this is not feasible in many scenarios, so the user input must be sanitized and validated (if possible). Traditional techniques use rules such as filters and regular expressions, which are not scalable and might not be able to capture more complex attack vectors. On the other hand, ML-based techniques learn directly from the data and have the potential to detect hidden patterns which would slip through traditional approaches. Indeed, several previous works reported high (99-100\%) detection rates, so ML-based methods are possible candidates for real-life deployment as they can aid security analysts with a measure of the maliciousness of an SQL payload. Below we give a high-level introduction to the data parsing techniques and ML architectures utilized in this work. 

\vspace{0.1cm}
\noindent\textbf{Pre-Processing. }
The raw benign and malicious SQL payloads cannot be fed directly into ML models. First, they must be pre-processed. We surveyed the relevant literature and identified three parsing techniques that are the most widely utilized: TF-IDF-based, Keyword based, and Skip-gram based. 

\begin{itemize}
	\item \textit{TF-IDF vectorizer} is based on the Bag of Words model that counts how much occurs from a word in a document. It consists of the Term Frequency (TF) and the Inverse Document Frequency (IDF) parts. TF measures the frequency of a word in a specific document, while IDF measures the importance of the words across the entire dataset. By combining these, those words will be more critical than rare words in the corpus or frequent ones in a specific document.

	\item \textit{Keyword weights} are based on the fact that SQL queries contain specific keywords that define how likely they are SQLi. We can assign weights to these keywords based on their maliciousness and extract features from the original text by using these weights.

	\item \textit{Skip-gram model} (opposite the previous two) considers semantics. It is a type of word embedding that means that every word is mapped into a continuous vector space, making it easier to check which ones are similar. A common way to create such word embeddings is to use the Word2Vec technique.
\end{itemize}

\vspace{0.1cm}
\noindent\textbf{Models. }
There are many model architectures choices to feed the processed data into. We surveyed the relevant literature and identified five ML architectures that are the most widely utilized: Linear Regression, Support Vector Machine, Random Forrest, Gradient Boosting, and Neural Networks.

\begin{itemize}
	\item \textit{Logistic Regression} (LR) is a simple linear model that learns to classify the data by minimizing the number of falsely categorized samples.

	\item \textit{Support Vector Machine} (SVM) learns to classify the data by maximizing the distance between the classes.
    
	\item \textit{Random Forest} (RF) consists of several Decision Trees that operate as an ensemble: the decision is based on the majority vote of the trees. 
 
	\item \textit{Gradient Boosting} (GB) is also an ensemble method; it is learning by minimizing the loss function, which is achieved by adding more weak learners, mainly concentrating on the areas where the already existing learners perform poorly.

	\item \textit{Neural Network} (NN) is a process that mimics how the human brain operates, i.e., it contains layers of neurons consisting of logistic units with an activation function. A wide multi-layer NN can capture more complex tasks than previously described models.
\end{itemize}

\section{Related Works}
\label{sec:3}

This section summarizes the previous research efforts in connection with SQLi detection using ML. Tackling this problem without ML is reviewed in \cite{kindy2011survey}, while \cite{pattewar2019detection} and \cite{hu2020survey} are surveyed the ML solutions. We inspected the ML-based SQLi literature\footnote{We considered 28 papers, which we obtained by forward and backward snowballing from the surveys and by using targeted queries (e.g., "SQL Injection" + "Machine Learning", etc.) in Google Scholar.} while focusing on four aspects: the dataset, the features, the models, and the evaluation. Our findings are summarized in Table \ref{tab:meta}. 

\begin{table*}[t]
    \begin{subtable}{\textwidth}
    \centering
    \resizebox{\textwidth}{!}{%
    \begin{tabular}{c|cccccccccccccc}
        R & \cite{joshi2014sql} & \cite{hasan2019detection} & \cite{moosa2010artificial} & \cite{chen2018research} & \cite{gandhi2021cnn} & \cite{pham2020experimental} & \cite{mishra2019sql} & \cite{krishnan2021sql} & \cite{ingre2017decision} & \cite{jothi2021efficient} & \cite{sheykhkanloo2015sql} & \cite{lou2019cnn} & \cite{yu2019detecting} & \cite{alam2021scamm} \\
        \hline
        D & $\cdot$ & $\cdot$ & $\cdot$ & $\cdot$ & $\cdot$ & $\cdot$ & $\cdot$ & $\cdot$ & $\circ$ & $\circ$ & $\circ$ & $\circ$ & $\circ$ & $\circ$ \\
        F & $\cdot$ & $\cdot$ & $\circ$ & $\circ$ & $\circ$ & $\circ$ & $\circ$ & $\circ$ & $\circ$ & $\circ$ & $\circ$ & $\bullet$ & $\bullet$ & $\bullet$ \\
        M & $\cdot$ & $\circ$ & $\cdot$ & $\cdot$ & $\circ$ & $\circ$ & $\bullet$ & $\bullet$ & $\cdot$ & $\cdot$ & $\cdot$ & $\cdot$ & $\cdot$ & $\circ$ \\
        E & $\circ$ & $\bullet$ & $\cdot$ & $\bullet$ & $\circ$ & $\circ$ & $\cdot$ & $\circ$ & $\circ$ & $\circ$ & $\circ$ & $\circ$ & $\circ$ & $\cdot$ \\
    \end{tabular}}\vspace{0.1cm}
    \resizebox{\textwidth}{!}{%
    \begin{tabular}{c|cccccccccccccc}
        R & \cite{chen2021sql} & \cite{ross2018sql} & \cite{uwagbole2017applied2} & \cite{li2019sql} & \cite{tripathy2020detecting} & \cite{tang2020detection} & \cite{hosam2021sql} & \cite{liu2020deepsqli} & \cite{farooq2021ensemble} & \cite{xie2019sql} & \cite{betarte2018web} & \cite{li2019lstm} & \cite{uwagbole2017applied} & \cite{gogoi2021defending}  \\
        \hline
        D & $\circ$ & $\circ$ & $\circ$ & $\bullet$ & $\bullet$ & $\bullet$ & $\bullet$ & $\bullet$ & $\bullet$ & $\bullet$ & $\bullet$ & $\bullet$ & $\bullet$ & $\bullet$  \\
        F & $\bullet$ & $\bullet$ & $\bullet$ & $\cdot$ & $\cdot$ & $\cdot$ & $\circ$ & $\circ$ & $\circ$ & $\bullet$ & $\bullet$ & $\bullet$ & $\bullet$ & $\bullet$  \\
        M & $\circ$ & $\circ$ & $\bullet$ & $\circ$ & $\circ$ & $\bullet$ & $\circ$ & $\circ$ & $\circ$ & $\cdot$ & $\circ$ & $\circ$ & $\circ$ & $\bullet$  \\
        E & $\circ$ & $\circ$ & $\bullet$ & $\circ$ & $\bullet$ & $\circ$ & $\cdot$ & $\cdot$ & $\bullet$ & $\bullet$ & $\circ$ & $\circ$ & $\circ$ & $\circ$  \\
    \end{tabular}}
    \caption{The symbol $\cdot$, $\circ$, and $\bullet$ means insufficient/medium/sufficient, as described in Table \ref{tab:note}. R, D, F, M, and E means \underline{R}eferences, \underline{D}ataset Size, Number of \underline{F}eatures, \underline{M}odel Optimization, and \underline{E}valuation Metrics respectively.}
    \label{tab:relatedwork}
    \end{subtable}\vspace{0.25cm}

    \begin{subtable}{\textwidth}
    \centering
    \resizebox{\textwidth}{!}{%
    \begin{tabular}{c||c|c|c|c}
         & \underline{D}ata & \underline{F}eature & \underline{M}odel & \underline{E}valuaton \\
        \hline
        $\cdot$ & $<10k$ & $<12$ & 1 w/o Tuning & acc. \\
        $\circ$ & $>10k$, $1$ source & $[12,999]$ & 1 w/ Tuning or $>1$ w/o Tuning & acc. \& conf. mx  \\
        $\bullet$ & $>10k$, $>1$ source & $>1000$ & $>1$ w/ Tuning & acc. \& conf. mx \& ROC \\
    \end{tabular}}
    \caption{Notation used in Table \ref{tab:relatedwork}. Acc. and conf. mx are the abbreviations for accuracy and confusion matrix.}
    \label{tab:note}
    \end{subtable}
    \caption{Comparision of varions SQLi detection works using ML.}
    \label{tab:meta}
\end{table*}

\vspace{0.1cm}
\noindent\textbf{Dataset. }
The datasets' size and diversity (i.e., distribution) are imperative concerning ML. Yet, more than a quarter (29\%) of works experiments with small (i.e., below 10k) datasets. Although the rest utilize a sufficient amount of data for training, for many of them (32\%), the data comes from a single source (aka distribution). Besides, when the authors consider multiple sources (39\%), they merge them to form an extensive central database split into training and testing. On the other hand, we train our models on many separate datasets from different sources and evaluate them in a cross-verification manner.

\vspace{0.1cm}
\noindent\textbf{Features. }
Few works (18\%) only utilize less than a dozen features, which is insufficient to capture the underlying language's richness. Although other works (43\%) exploit more features, only some of them (39\%) apply over a thousand features (i.e., by using OneHot-Encoding, Word2Vec, String2Vec, or TF-IDF with large datasets), which is the best practice in NLP and needed to capture the abundance of the payloads appropriately. 

\vspace{0.1cm}
\noindent\textbf{Models. }
Almost a third of the works (32\%) mentioned in Table \ref{tab:meta} consider only a single ML model without any hyper-parameter tuning. This cherry-picking strategy is superficial and, without proper comparison, could be easily misinterpreted. Although other works (57\%) consider comparing more off-the-shelf models or fine-tuning a single one, this still not paints a complete picture of the relationship between these models. Finally, similarly to our work, only a handful of papers (18\%) evaluate multiple models and utilize parameter optimization. 

\vspace{0.1cm}
\noindent\textbf{Evaluation. }
Few works (18\%) present the accuracy metric only, which is inappropriate in the SQLi use-case: the difference between type I and type II errors is crucial. The majority of the works (61\%) indeed consider false positives and false negatives and present them either via the confusion matrix or via the precision, recall, and F1 values. However, this still might be insufficient from the usability point of view: any practitioner of an SQLi detection system would require the possibility to set the trade-off between these values, depending on the underlying scenario's sensitivity. Hence, the ROC curve is of the utmost importance. Besides this work, it is measured only half a dozen times (21\%).

\subsection*{Discussion}

Consequently, in Table \ref{tab:relatedwork}, we can see that no work in the existing literature thoroughly tackles all four aspects we considered. Indeed, only eight works (29\%) consider all four aspects to some extent (i.e., no $\cdot$ sign in a column), and only three (11\%) consider three aspects sufficiently (i.e., three $\bullet$ sign in a column). 

Moreover, none of these publications are from highly regarded security venues. In fact, most (86\%) were published at places not dealing directly with security, and some (39\%) even appeared outside the computer science domain. Hence, research focusing on SQLi detection using ML could be more satisfactory, and many existing works should be taken with a grain of salt.

\section{Experiments}
\label{sec:4}

\begin{figure*}[t]
    \centering
    \includegraphics[width=12cm]{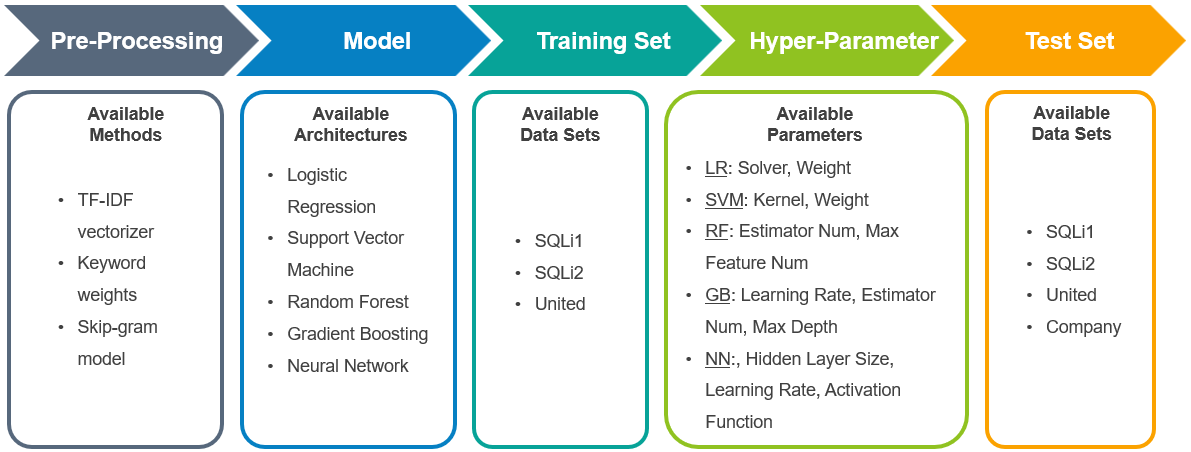}
    \caption{The Process Diagram for our experiments. Various pre-processing (1) and model architectural (2) choices are available, the training set (3) and the testing set (5) could also be varied, moreover, for each scenario an appropriate hyper-parameter tuning (4) is included.}
    \label{fig:dia}
\end{figure*}

In this section, we present our experimental setups and the corresponding results. To honor blind submission, we share our implementation only after acceptance. Figure \ref{fig:dia} presents a summarizing process diagram. 

\vspace{0.1cm}
\noindent\textbf{Datasets. }
Besides providing a comprehensive analysis, our main aim is to compare models on different datasets with various sizes coming from distinct distributions. Thus, obtaining appropriate datasets is crucial. For our experiments, we utilized three public datasets with different sizes (small, medium, large) from two sources. We merged three small datasets (OWASP, BurpSuite, and FuzzDB) from GitHub\footnote{\url{https://www.github.com/ChrisAHolland/ML-SQL-Injection-Detector/tree/master/data}} into one we called \textit{United}. We also used two datasets from Kaggle\footnote{\url{https://www.kaggle.com/datasets/syedsaqlainhussain/sql-injection-dataset}}, namely \textit{SQLi1} and \textit{SQLi2}. Finally, we employed a private dataset only for testing (referred to as \textit{Company} from a SIEM of an international SOC operating company with clients all over Europe. Opposed to the first three datasets, the last one is not public. The data belongs to a single client, and it was acquired in-between 2019/08 and 2021/05. The utilized datasets are summarized in Table \ref{tab:data}.

 \begin{table}[!b]
    \centering
    \begin{tabular}{c|ccc}
        Name & Size & Benign & Malicious \\
        \hline
        United & 1140 & 1133 & 7 \\
        SQLi1 & 3950 & 950 & 3000 \\
        SQLi2 & 33725 & 11424 & 22301 \\
        Company & 2594 & 2337 & 257 \\
    \end{tabular}
    \caption{The properties of the considered datasets. }
    \label{tab:data}
\end{table}

 \vspace{0.1cm}
\noindent\textbf{Hyper-Parameter Tuning. }
For each model type, we fine-tuned several hyper-parameters to obtain the best results. Note that our goal is not to produce state-of-the-art performing models but to show that optimizing various models for one distribution affects these models differently when tested on other distributions. We used sklearn\footnote{\url{https://www.scikit-learn.org/stable}} for our implementation. We performed a grid search on the following parameters while the rest we set to be the default. 

\begin{itemize}
    \item \textit{Logistic Regression}: the two parameters we fine-tune are 1) the type of the \textit{\textbf{S}olver} ($\mathtt{newton-cg}, \mathtt{lbfgs}$) and 2) the \textit{\textbf{W}eight} of regularization ($10^{-1},10^{0},10^{1}$).
    \item \textit{Support Vector Machine}: the two parameters we fine-tune are 1) the \textit{\textbf{K}ernel} type ($\mathtt{Linear}, \mathtt{Polynomial}, \mathtt{Rbf}$) and 2) the \textit{\textbf{W}eight} of regularization ($10^{-1},10^{0},10^{1}$).
    \item \textit{Random Forest}: the two parameters we fine-tuned are 1) the max number of \textit{\textbf{F}eatures} ($2^0,2^1,\dots,2^5$) and 2) the number of \textit{\textbf{E}stimators} ($10^{1},10^{2},10^{3}$).
    \item \textit{Gradient Boosting}: the three parameters we fine-tuned are 1) the \textit{\textbf{L}earning Rate} ($10^{-3},10^{-2},10^{-1}$), 2) the number of \textit{\textbf{E}stimators} ($10^{1},10^{2},10^{3}$), and 3) the maximum \textit{\textbf{D}epth} ($2, 4, 8$).
    \item \textit{Neural Network}: the three parameters we fin-tuned are 1) the \textit{\textbf{L}earning Rate} ($10^{-3},10^{-2},10^{-1}$), 2) the size of the \textit{\textbf{H}idden Layer} ($64, 128$), and 3) the \textit{\textbf{A}ctivation Function} ($\mathtt{ReLU}, \mathtt{Sigmoid}$). 
\end{itemize}

\begin{table*}[t]
    \centering
    \resizebox{\textwidth}{!}{%
    \begin{tabular}{|c|c|c|c|c|c|c|}
        \hline
        DataSet & Pre-Proc. & Model & Parameters & Train & Val. & Test\\
        \hline
        \multirow{5}{*}{United} & Skip-gram & LR & S:$\mathtt{newton}$,W:0.1 & 99.69\% & 99.56\% & 99.78\%  \\
        & Skip-gram & SVM & K:$\mathtt{linear}$,W:10 & 99.94\% & 100\% & 100\%  \\
        & Skip-gram & RF & F:32,E:10 & 100\% & 100\% & 99.89\%  \\
        & Skip-gram & GB & L:0.01,E:1000,D:2 & 100\% & 100\% & 99.78\%  \\
        & Skip-gram & NN & L:0.001,H:64.A:$\mathtt{sigmoid}$ & 99.69\% &99.56\% & 99.78\%  \\
        \hline
        \multirow{5}{*}{SQLi1} & TF-IDF & LR & S:$\mathtt{newton}$,W:10 & 98.73\% & 96.77\% & 98.43\%  \\
        & TF-IDF & SVM & K:$\mathtt{linear}$,W:1 & 98.20\% & 96.34\% & 97.86\%   \\
        & TF-IDF & RF & F:16,E:100 & 100\% & 97.16\% & 97.79\%  \\
        & TF-IDF & GB & L:0.1,E:1000,D:4 & 100\% & 98.70\% & 98.68\%  \\
        & TF-IDF & NN & L:0.001,H:64,A:$\mathtt{sigmoid}$ & 93.88\% & 90.77\% & 92.78\%  \\
        \hline
        \multirow{5}{*}{SQLi2} & TF-IDF & LR & S:$\mathtt{newton}$,W:10 & 99.54\% & 99.25\% & 99.36\%  \\
        & Skip-gram & SVM & K:$\mathtt{poly}$, W:10 & 99.32\% & 99.47\% & 99.41\%  \\
        & Skip-gram & RF & F:1,E:100 & 99.99\% & 99.61\% & 99.57\%  \\
        & Skip-gram & GB &  L:0.1,E:1000,D:8 & 100\% & 99.34\% & 99.30\%  \\
        & TF-IDF & NN & L:0.1,H:64,A:$\mathtt{sigmoid}$ & 99.56\% & 99.45\% & 99.34\%  \\
        \hline
    \end{tabular}}
    \caption{For all considered public datasets and models, we present the F1-scores of the best-performing models with the corresponding pre-processing methods for the training set, the validation set, and the test set. The utilized hyper-parameters are also displayed where S, W, K, F, E, L, D, H, and A are Solver, Weight, Kernel, Feature num., Estimator num., Learning rate, Depth, Hidden layer size, and Activation function, respectively. }
    \label{tab:exp1}
\end{table*}

\subsection{Results}

We considered three scenarios to evaluate. Firstly, to give a comprehensive analysis, we review the well-studies IID case (i.e., when the test and train datasets are from the same distribution). Secondly, to measure the robustness of the models against data distribution change, we provide experiments concerning the non-IID case (which was not studied before), namely when the training and the testing data come from a different distribution. Thirdly, to inspect the applicability of the lab-tested models in the real world, we evaluate the trained models on confidential data of an international SOC operator within Europe. When applicable, we randomly split the datasets into training, validation, and testing using 70-10-20 percentages. All our experiments are performed two-fold to mitigate the randomness of the training process. 

\begin{figure*}[t]
    \centering
    \includegraphics[width=12cm]{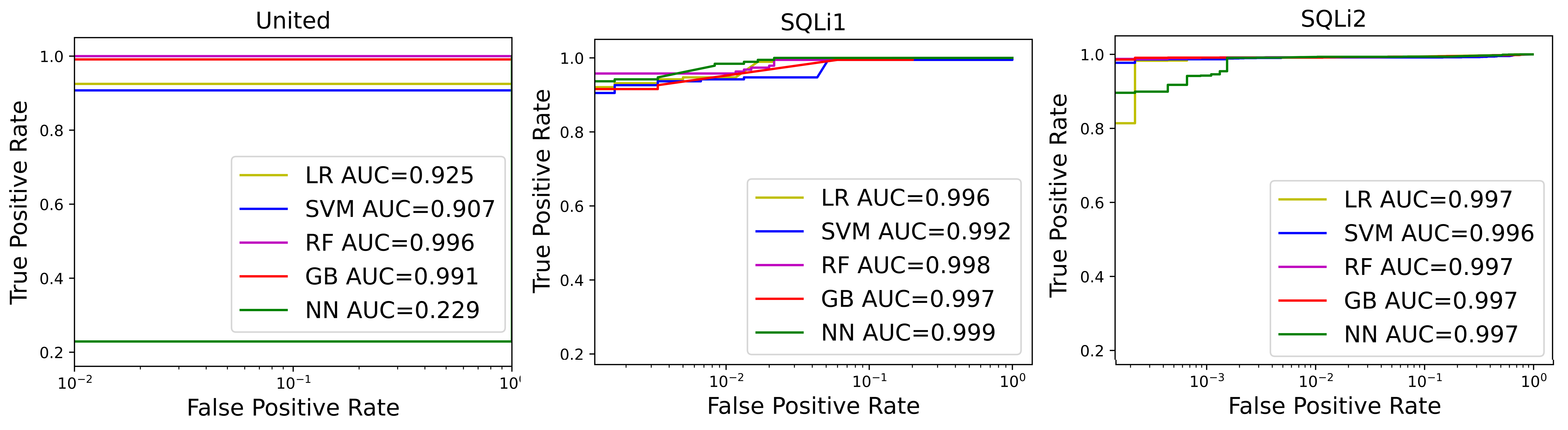}
    \caption{The best performing setting's ROC curves with the AUC values for all datasets.}
    \label{fig:exp1_roc}
\end{figure*}

\vspace{0.1cm}
\noindent\textbf{Using the same distribution for Training/Testing. }
Due to the lack of space, we neither show the accuracy nor the confusion matrices but instead present the more informative F1-scores and the ROC curves. The former is visible in Table \ref{tab:exp1}, while the latter is visualized in Figure \ref{fig:exp1_roc} for the considered three datasets (United, SQLi1, SQLi2). In  Table \ref{tab:exp1}, we also present the best-performing model types (LR, SVM, RF, GB, NN) with the corresponding optimal pre-processing method (TF-IDF, Keyword, Skip-gram) and hyper-parameters. The models trained on United (with Skip-gram) have 125 features, the models trained on SQLi1 (with TF-IDF) have 9683, and the models trained on SQLi2 have 1455 and 28679 when pre-processed with Skip-gram and TF-IDF respectively.

In Table \ref{tab:exp1}, one can see that the best-performing setups (pre-processing, model type, hyper-parameters) across different datasets vary greatly. For instance, Skip-gram pre-processing method outperforms TF-IDF on the United dataset, while the opposite trend corresponds to SQLi1, and neither dominates the other on SQLi2. The optimal learning rate for GB and NN and the optimal weight for LR and SVM depend on the underlying dataset. No one model type dominates, i.e., the model obtaining the highest F1-score is different for all three datasets. Hence, it is uttermost important to see how models optimized for one dataset perform on other datasets with different distributions.

In Figure \ref{fig:exp1_roc} from the ROC curves, it is visible that independently of the optimal setup (i.e., model type, pre-processing method, hyper-parameters), RF slightly outperforms the other models in the low false positive rate region as it obtains the highest true positive rate. Conversely, when a high false positive rate is tolerated, the models have only negligible differences. Note that the AUC values are all above 0.99 except for the United dataset due to its small size: there is only a single negative sample in its test set. 

In addition to these results, we found that the Keyword weights pre-processing method is inferior to both TF-IDF and Skip-gram, as the corresponding results were always about 10\% less, even though besides the model parameters, we also tuned the exact weights for this pre-processing method. 

\vspace{0.1cm}
\noindent\textbf{Timing Measurements. }
Besides its prediction power, another essential aspect is the usability of the models, i.e., how much time it takes to train these models, what is their sizes, and how fast they can predict. These details are presented in Table \ref{tab:speed} for the best-performing models. The training was done on Ubuntu 20.04.4 LTS Linux with 16 CPUs (3.10GHz) and 98 Gb RAM. One can see that while the models achieve comparable performances, both the time and the size values have a considerable variance. Additionally to the model type and the employed hyper-parameters, these differences are a combined result of the corresponding datasets and pre-processing methods. Yet, two trends are visible: LR is always the smallest model, and GB is always the most costly model to be trained. Along with the ROC curve, such information is essential for SOC operators to optimize the trade-off between the usability and prediction performance of the SQLi-detecting ML model.

\begin{table}[t]
    \centering
    \begin{tabular}{|c|c|c|c|c|}
        \hline
        Dataset & Mod. \& PreP. & Learn. Time & Mod. Size & Pred. Speed \\
        \hline
        \multirow{5}{*}{United} & LR (S) & 0.0258 s & 0.002 Mb & 0.002 ms \\
        & SVM (S) & 0.0204 s & 0.023 Mb & 0.002 ms \\
        & RF  (S) & 0.018 s & 0.012 Mb & 0.001 ms \\
        & GB  (S) & 1.4178 s & 0.646 Mb & 0.005 ms \\
        & NN  (S) & 0.283 s & 0.118 Mb & 0.054 ms \\
        \hline
        \multirow{5}{*}{SQLi1} & LR (T) & 1.9164 s & 0.219 Mb & 0.123 ms \\
        & SVM (T) & 44.1636 s & 72 Mb & 3.077 ms \\
        & RF  (T) & 2.3232 s & 7.1 Mb & 0.174 ms \\
        & GB  (T) & 589.8 s & 1.5 Mb & 0.150 ms \\
        & NN  (T) & 1.5184 s & 7.2 Mb & 0.112 ms \\
        \hline
        \multirow{5}{*}{SQLi2} & LR (T) & 57.7254 s & 0.631 Mb & 0.320 ms \\
        & SVM (S) & 35.2536 s & 3.4 Mb & 0.089 ms \\
        & RF  (S) & 3.5634 s & 4.2 Mb & 0.043 ms \\
        & GB  (S) & 572.54 s & 6.8 Mb & 0.035 ms \\
        & NN  (T) & 21.4831 s & 21 Mb & 0.165 ms \\
        \hline
    \end{tabular}
    \caption{The pre-processing and training time, the model size, and the prediction speed of the best performing models with \textbf{S}kip-gram or with \textbf{T}F-IDF.}
    \label{tab:speed}
\end{table}

\begin{figure*}[t]
    \centering
    \includegraphics[width=12cm]{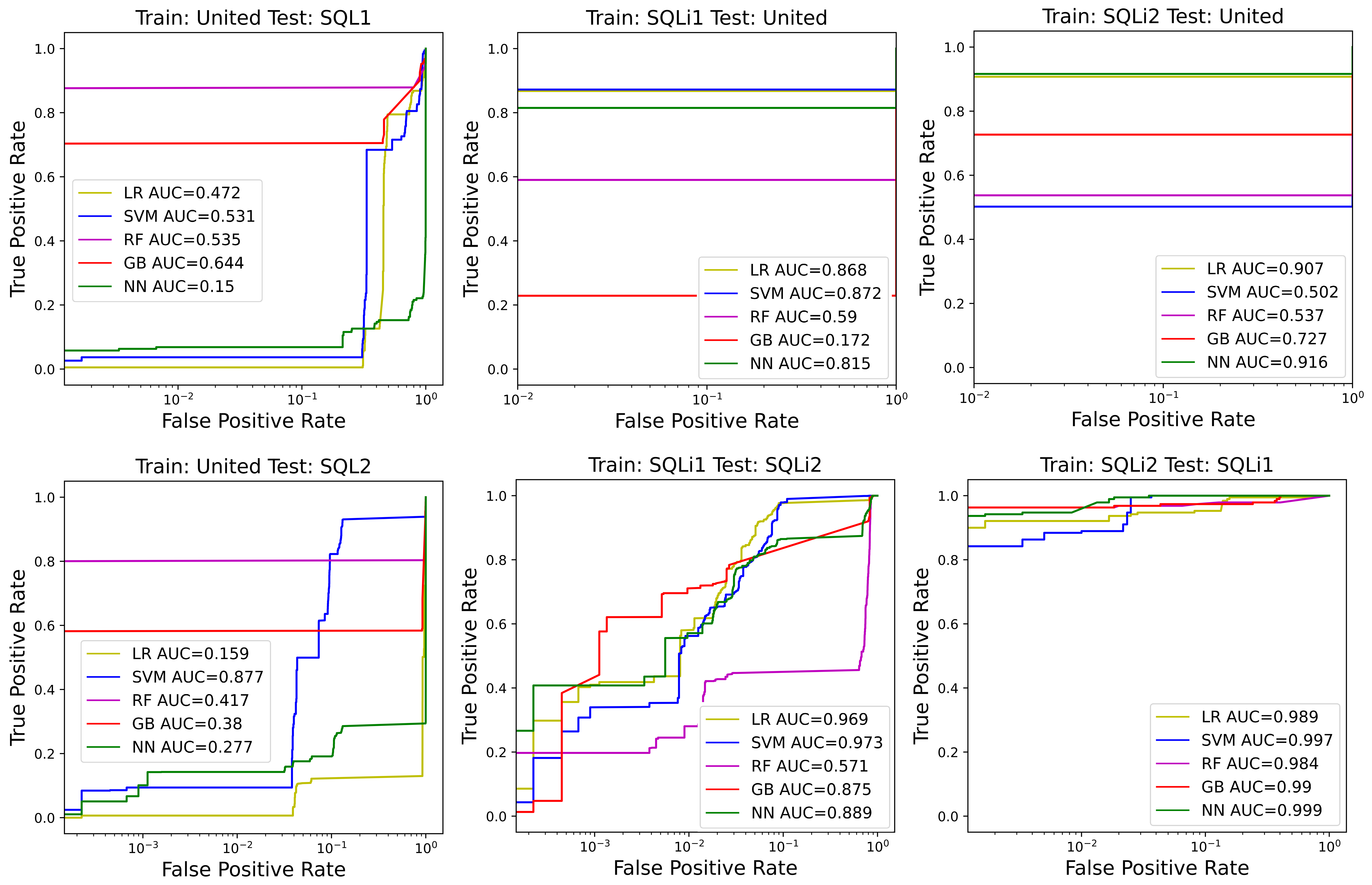}
    \caption{The best performing setting's ROC curves with the AUC values for all cross-verification settings.}
    \label{fig:exp2_roc}
\end{figure*}

\vspace{0.1cm}
\noindent\textbf{Using different distribution for Training/Testing. }
The previous experiments revealed the sensitivity of the setup: similar high F1-scores could be reached with vastly different settings. Opposed to the common IID practice that uses the same distribution for testing and training (by splitting the same dataset), we are focusing on the non-IID case, i.e., measuring the performance of the models on test sets from other distributions. This experiment measures the model's robustness against data distribution. The F1-scores are shown in Table \ref{tab:exp2}, while the ROC curves for all pair-wise scenarios are presented in Figure \ref{fig:exp2_roc}.

As expected, the F1-score of the best-performing models drops when tested on other datasets from other distributions. For instance, training on a small dataset could produce completely unreliable models: the result on the top of Table \ref{tab:exp2} suggests that the models trained on United are essentially reduced to a random guess when tested on SQLi1 and SQLi2. Similar results can be found on the left side of Figure \ref{fig:exp2_roc}: when trained on the smallest United dataset, the best models' AUC is  0.644 and 0.877 when tested on SQLi1 and SQLi2, respectively. 

Additionally, the pre-processing method seems crucial too: when trained on SQLi2 and tested on United, models using Skip-gram are idle. In contrast, the ones using TF-IDF have a decent performance. Another interesting finding is that different models could have opposing generalization properties against other distributions. For example, in the middle of Table \ref{tab:exp2} RF performs exceptionally on United and terribly on SQLi2 when trained on SQLi1. At the same time, the exact opposite trend holds for GB. 

\begin{table}[t]
    \centering
    \begin{tabular}{|c||c|c|c|c|c|}
        \hline
        Tested & \multicolumn{5}{c|}{Trained on United} \\
        \cline{2-6}
        on & LR (S) & SVM (S) & RF (S) & GB (S) & NN (S) \\
        \hline
        United & 99.78\% & 100\% & 99.89\% & 99.78\% & 99.78\% \\
        SQLi1 & 38.8\% & 39.0\% & 38.8\% & 37.8\% & 38.8\% \\
        SQLi2 & 50.6\% & 49.6\% & 50.6\% & 50.5\% & 50.6\% \\
        \hline
    \end{tabular}
    \vspace{0.1cm}
    \begin{tabular}{|c|c|c|c|c|c|}
        \hline
        Tested & \multicolumn{5}{c|}{Trained on SQLi1} \\
        \cline{2-6}
        on & LR (T) & SVM (T) & RF (T) & GB (T) & NN (T) \\
        \hline
        United & 92.2\% & 92.4\% & 99.8\% & 76.1\% & 91.2\% \\
        SQLi1 & 98.43\% & 97.86\% & 97.79\% & 98.68\% & 92.78\% \\
        SQLi2 & 84.6\% & 79.7\% & 52.8\% & 82.2\% & 83.2\% \\
        \hline
    \end{tabular}
    \vspace{0.1cm}
    \begin{tabular}{|c|c|c|c|c|c|}
        \hline
        Tested & \multicolumn{5}{c|}{Trained on SQLi2} \\
        \cline{2-6}
        on & LR (T) & SVM (S) & RF (S) & GB (S) & NN (T) \\
        \hline
        United & 97.3\% & 41.8\% & 50.7\% & 66.5\% & 97.5\% \\
        SQLi1 & 95.3\% & 92.5\% & 97.8\% & 98.1\% & 96.5\% \\
        SQLi2 & 99.36\% & 99.41\% & 99.57\% & 99.30\% & 99.34\% \\
        \hline
    \end{tabular}
    \caption{Dataset-wise the F1-scores with cross-verification of the best performing models with \textbf{S}kip-gram or with \textbf{T}F-IDF.}
    \label{tab:exp2}
\end{table}

Concerning the ROC curves in Figure \ref{fig:exp2_roc}, similarly to the IID case, for this non-IID setup RF is also ideal for low false positive rate region but only when trained on United (left). However, when the models are trained on the large SQLi2 dataset (right), the highest AUC belongs to NN: 0.916 and 0.999 when tested on United and SQLi1, respectively. NN is also a good choice when a low false positive rate is desired. 

In addition, when the models are trained on SQLi1 and tested on United (i.e., middle top), SVM is the optimal model choice for sensitive domains where a low false positive rate is required. Yet, when tested on SQLi2 (i.e., middle bottom), multiple models achieve the best trade-off, depending on the desired false positive rate range. What is clear is SVM has the highest AUC values (0.872 and 0.973).

Based on these results, TF-IDF and SVM seem more robust against test data distribution shifts than other methods and models. TF-IDF uses statistical features such as frequencies, which change only slightly when there is a minor change in the underlying distribution. On the other hand, Skip-gram is based on a Neural Network, which takes a predefined input size and could easily overfit, making it rigid to use with different input families. Considering models, SVM is robust, as it maximizes the smallest distance between the benign and malicious samples. Thus it should be tolerant of minor changes in the classes. GB also performed well in this experiment due to its assembly nature. In contrast, NN, the most complex model, might overfit (when trained on United) and lose its generalization capability to tackle samples from other distributions. However, it could also outperform the rest of the models when trained on a large representative dataset, e.g., SQLi2. 

\begin{figure*}[t]
    \centering
    \includegraphics[width=12cm]{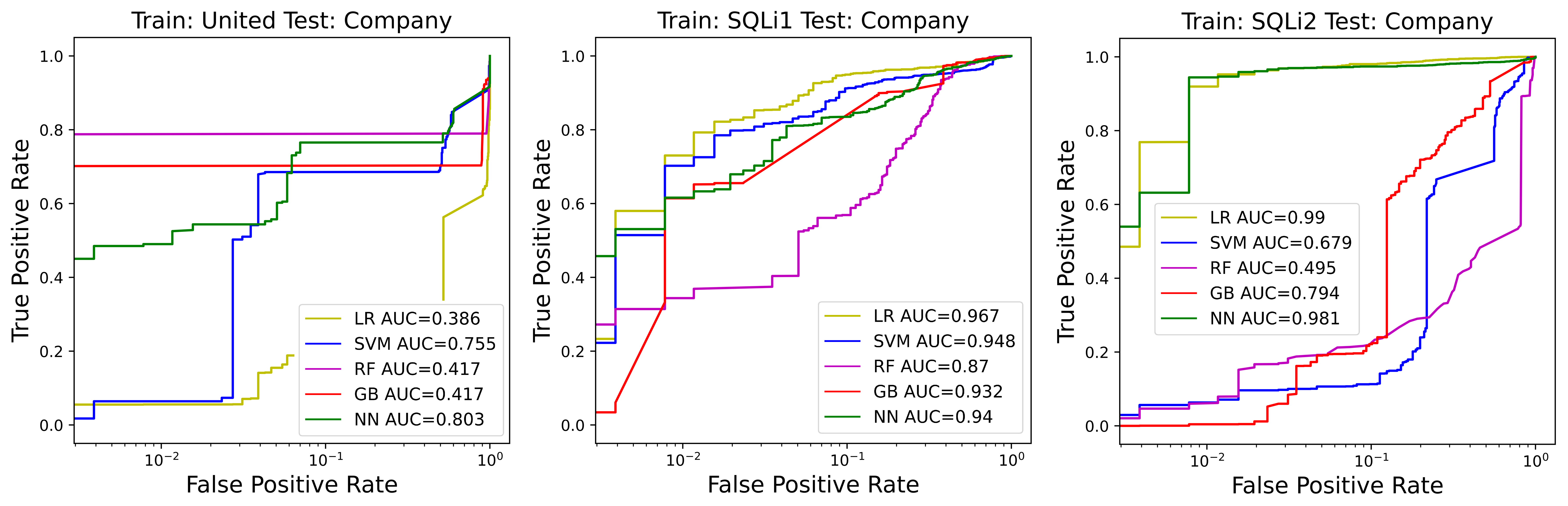}
    \caption{The best performing setting's ROC curves with the AUC values trained on the entire public domain and tested on a private dataset.}
    \label{fig:exp3_roc}
\end{figure*}

\vspace{0.1cm}
\noindent\textbf{Validating the Findings on Private dataset. }
Finally, we perform a similar non-IID experiment, but instead of utilizing public datasets for training and testing, we apply the private Company dataset as a test set. The F1-scores of the best performing models are shown in Table \ref{tab:exp3} while the ROC curves are presented in Figure \ref{fig:exp3_roc}.

The last column of Table \ref{tab:exp3} (using the large SQLi2 dataset) elaborates that Skip-gram is indeed not appropriate for models indented to be used on other distributions than the model was trained on. This is seemingly contradicted in the first column; however, that corresponds to the smallest United dataset, which could also produce highly unreliable models, as we showed in Table \ref{tab:exp2}. We hypothesize this excellent result is due to the closeness of United's and Company's distribution. Similarly to the previous use cases, the highest F1-score (95.7\%) is reached by NN when trained on the biggest dataset using the robust TF-IDF.

Surprisingly, Figure \ref{fig:exp3_roc} (right) shows the simple LR model trained on SQLi2 does outperform NN based on the ROC with a minor AUC difference (0.99 vs. 0.98). Furthermore, LR also performs exceptionally on Company when trained on SQLi1 (middle), making it the best choice even for the low false positive rate domain. Finally, contrary to what the F1-scores suggest when trained on United, the AUC values are not exceptional. 

\begin{table}[!b]
    \centering
    \begin{tabular}{c|ccc}
    Trained on & United & SQLi1 & SQLi2 \\
    \hline\hline
    Model & \multicolumn{3}{c}{F1-score on Company} \\
    \hline
    LR  & 94.79\% (S) & 79.08\% (T) & 92.35\% (T) \\
    SVM & 90.97\% (S) & 77.88\% (T) & 29.98\% (S) \\
    RF  & 92.47\% (S) & 90.03\% (T) & 32.68\% (S) \\
    GB  & 91.69\% (S) & 76.41\% (T) & 78.47\% (S) \\
    NN  & 94.79\% (S) & 77.65\% (T) & 95.69\% (T) \\
    \end{tabular}
    \caption{F1-scores of the best performing models with \textbf{S}kip-gram or with \textbf{T}F-IDF when tested on the private Company data.}
    \label{tab:exp3}
\end{table}

\vspace{0.1cm}
\noindent\textbf{Discussion. }
The 'one size fits all' ideology needs to be revised in connection with ML models trained for SQLi detection. Our experiments uncovered that none of the setups is adequate for all environments with different constraints. Still, several rules of thumb could be identified via the presented results. Below we summarize our suggestions for several use cases. 

\begin{itemize}
    \item Limited resources \& similar usage: The smallest and fastest model from the studied models is LR. Besides, its performance is also decent, making it an optimal choice for IoT scenarios.
    
    \item Much resources \& limited datapoints \& different usage: Although SVM might need more resources than other models (in terms of model size and prediction speed), with TF-IDF it is the appropriate choice, as it is capable of separating benign and malicious samples even from other distributions with high accuracy when only a few samples are available for training. 
    
    \item Limited resources \& limited data points \& different usage: Since GB consumes too much resource, the fitting assembly model for this scenario is RF. 
    
    \item Many resources \& enough data points: As expected, without environmental constraints and enough data available, the most complex model (NN) with TF-IDF does outperform the rest. If not overfitted, NN is also appropriate for other distributions. 
\end{itemize}

\section{Conclusion}
\label{sec:5}

SQL Injections are top-tier vulnerabilities of today's ICT systems. As with many other problems, machine learning techniques have also been proven appropriate to tackle this issue. In this work, we highlighted the shortcomings of the previous machine learning solutions, which consider only a few aspects of the underlying problem. Thus, this study is the first to provide a comprehensive (wide and in-depth) empirical analysis of SQL injection detection via machine learning. Furthermore, we cross-validated the trained models by using data from other distributions. This aspect is idle in the literature, even though the sensitivity of models to distribution change is crucial for any real-life deployment. Our work could be beneficial for security engineers and practitioners working with SQL. 

\bibliographystyle{alpha}
\bibliography{ref.bib}

\end{document}